\documentclass[reprint, superscriptaddress, twocolumn, amsmath, amssymb, aps, prb]{revtex4-2}
\usepackage{graphicx}
\usepackage{dcolumn}
\usepackage{bm}
\usepackage{placeins}
\usepackage{appendix}

\usepackage{verbatim}
\usepackage[usenames,dvipsnames]{xcolor}
 \usepackage{amsmath}
 \usepackage{amsfonts}
 \usepackage{amssymb}
 \usepackage[colorlinks=true,citecolor=blue,linkcolor=red]{hyperref}

 \usepackage{bbold}     
 \usepackage[makeroom]{cancel}  
 \usepackage{multirow}    
 \usepackage[normalem]{ulem}        
 \usepackage{array}
 \usepackage{pdfpages}
\makeatletter
 \AtBeginDocument{\let\LS@rot\@undefined}
 \makeatother

\begin{document}


\title{Plateau Regions for Zero-Bias Peaks within 5$\%$ of the Quantized Conductance Value $2e^2/h$}


\author{Zhaoyu Wang}
 \email{equal contribution}
\affiliation{State Key Laboratory of Low Dimensional Quantum Physics, Department of Physics, Tsinghua University, Beijing 100084, China}

\author{Huading Song}
\email{equal contribution}
\email{songhd@baqis.ac.cn}
\affiliation{Beijing Academy of Quantum Information Sciences, 100193 Beijing, China}

\author{Dong Pan}
 \email{equal contribution}
\affiliation{State Key Laboratory of Superlattices and Microstructures, Institute of Semiconductors, Chinese Academy of Sciences, P. O. Box 912, Beijing 100083, China}

\author{Zitong Zhang}
\email{equal contribution}
\affiliation{State Key Laboratory of Low Dimensional Quantum Physics, Department of Physics, Tsinghua University, Beijing 100084, China}

\author{Wentao Miao}
\affiliation{State Key Laboratory of Low Dimensional Quantum Physics, Department of Physics, Tsinghua University, Beijing 100084, China}

\author{Ruidong Li}
\affiliation{State Key Laboratory of Low Dimensional Quantum Physics, Department of Physics, Tsinghua University, Beijing 100084, China}

\author{Zhan Cao}
\affiliation{Beijing Academy of Quantum Information Sciences, 100193 Beijing, China}

\author{Gu Zhang}
\affiliation{Beijing Academy of Quantum Information Sciences, 100193 Beijing, China}

\author{Lei Liu}
\affiliation{State Key Laboratory of Superlattices and Microstructures, Institute of Semiconductors, Chinese Academy of Sciences, P. O. Box 912, Beijing 100083, China}

\author{Lianjun Wen}
\affiliation{State Key Laboratory of Superlattices and Microstructures, Institute of Semiconductors, Chinese Academy of Sciences, P. O. Box 912, Beijing 100083, China}

\author{Ran Zhuo}
\affiliation{State Key Laboratory of Superlattices and Microstructures, Institute of Semiconductors, Chinese Academy of Sciences, P. O. Box 912, Beijing 100083, China}

\author{Dong E. Liu}
\affiliation{State Key Laboratory of Low Dimensional Quantum Physics, Department of Physics, Tsinghua University, Beijing 100084, China}
\affiliation{Beijing Academy of Quantum Information Sciences, 100193 Beijing, China}
\affiliation{Frontier Science Center for Quantum Information, 100084 Beijing, China}

\author{Ke He}
\affiliation{State Key Laboratory of Low Dimensional Quantum Physics, Department of Physics, Tsinghua University, Beijing 100084, China}
\affiliation{Beijing Academy of Quantum Information Sciences, 100193 Beijing, China}
\affiliation{Frontier Science Center for Quantum Information, 100084 Beijing, China}

\author{Runan Shang}
\affiliation{Beijing Academy of Quantum Information Sciences, 100193 Beijing, China}

\author{Jianhua Zhao}
 \email{jhzhao@semi.ac.cn}
\affiliation{State Key Laboratory of Superlattices and Microstructures, Institute of Semiconductors, Chinese Academy of Sciences, P. O. Box 912, Beijing 100083, China}

\author{Hao Zhang}
\email{hzquantum@mail.tsinghua.edu.cn}
\affiliation{State Key Laboratory of Low Dimensional Quantum Physics, Department of Physics, Tsinghua University, Beijing 100084, China}
\affiliation{Beijing Academy of Quantum Information Sciences, 100193 Beijing, China}
\affiliation{Frontier Science Center for Quantum Information, 100084 Beijing, China}


\begin{abstract}

Probing an isolated Majorana zero mode is predicted to reveal a tunneling conductance quantized at $2e^2/h$ at zero temperature.  Experimentally, a zero-bias peak (ZBP) is expected and its height should remain robust against relevant parameter tuning, forming a quantized plateau. Here, we report the observation of large ZBPs in a thin InAs-Al hybrid nanowire device. The ZBP height can stick close to $2e^2/h$, mostly within $5\%$ tolerance, by sweeping gate voltages and magnetic field. We further map out the phase diagram and identify two plateau regions in the phase space. Despite the presence of disorder and quantum dots, our result constitutes a step forward towards establishing Majorana zero modes.

\end{abstract}

\maketitle

Majorana zero modes (MZMs) \cite{ReadGreen,Kitaev} have been extensively searched in hybrid semiconductor-superconductor nanowire devices since the first material prediction in 2010 \cite{Lutchyn2010,Oreg2010}. One key prediction, a quantized zero bias peak (ZBP) in tunneling conductance \cite{DasSarma2001, Law2009, Flensberg2010, Wimmer2011QPC}, still remains illusive so far \cite{Mourik, Deng2016, Nichele2017, Gul2018, Zhang2021, Song2021, Prada2020, NextSteps}. Moreover, theory developments have proposed the concept of quasi-MZMs due to smooth potential variation \cite{BrouwerSmooth, Prada2012, WimmerQuasi, TudorQuasi,Aguado_quasi_MZM, CaoZhanPRL,Aguado_non_hermitian} or disorder \cite{GoodBadUgly, DasSarma2021Disorder, Tudor2021Disorder,SauQuality, on-demand, DasSarma_random_matrix}. These quasi-MZMs, though topologically trivial, can also lead to quantized ZBPs. The quantization mechanisms for MZMs and quasi-MZMs are similar: the conductance is solely contributed by one isolated MZM while the second MZM is decoupled. Recent experimental progress has reported large ZBPs whose height can reach $2e^2/h$ \cite{Zhang2021, Song2021}. However, a single peak at $2e^2/h$ is not enough to be entitled as `quantized'. A quantized ZBP requires the peak height sticking to $2e^2/h$ by tuning all relevant experimental knobs: a plateau defined by parameter sweepings. This plateau phenomenon is still missing in experiments. 

Here, we have improved the device fabrication and report ZBPs near $2e^2/h$, forming a plateau defined by sweeping gate voltages and magnetic field ($B$) in an ultra-thin InAs-Al nanowire device. We quantify the plateau with a tolerance of $5\%$, a number commonly used in recent literature \cite{Tudor2021Disorder, SauQuality}. In the end, we discuss their possible connections to MZMs or quasi-MZMs.

\begin{figure}[htb]
\includegraphics[width=\columnwidth]{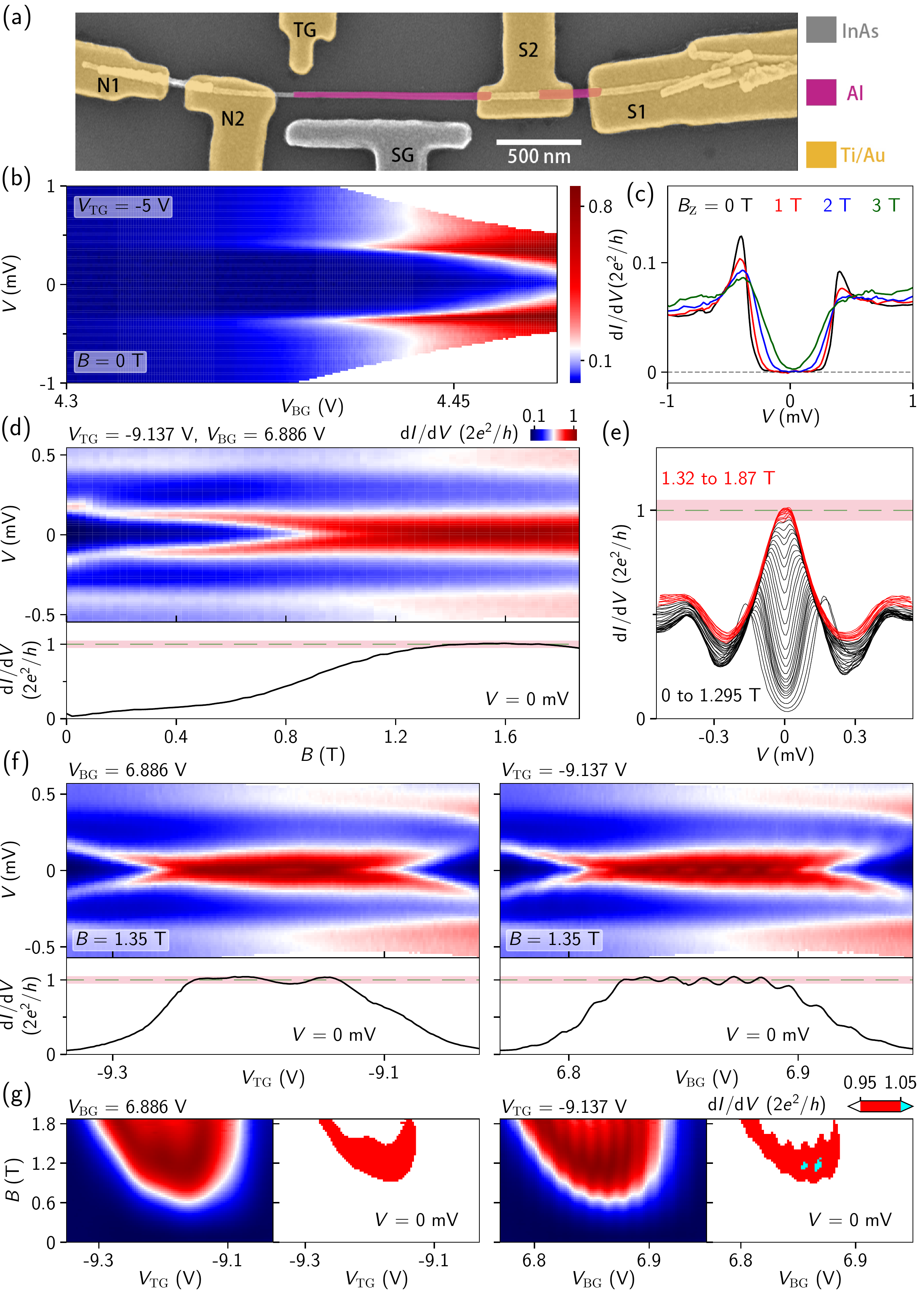}
\centering
\caption{(a) Device SEM (false colored). The contacts and gates are Ti/Au (5/70 nm in thickness). The substrate is p-doped Si covered by 300 nm thick SiO$_2$, serving as a global back gate (BG). (b) Hard gap tunneling spectroscopy at 0 T. (c) $B_{\text{z}}$ dependence of the gap. (d) $B$ scan (aligned with the nanowire) of a ZBP near $2e^2/h$. Lower panel, zero-bias line cut. The pink bar marks the range from 0.95 to 1.05 in the unit of $2e^2/h$ for all figures. (e) `Waterfall' plots of (d). For clarity, every other line cut is shown and two colors are used. (f) $V_{\text{TG}}$ and $V_{\text{BG}}$ scans of the ZBP. (g) Gate and $B$ scans at zero bias. The three-color plots mark red as regions within $\pm 5 \%$ of $2e^2/h$. (d) (f) and (g) share the same color bar. }
\label{fig1}
\end{figure}

Figure 1a shows the device scanning electron micrograph (SEM). Material and growth details can be found in Ref \cite{PanCPL}. The InAs diameter ($\sim$ 26 nm) is much thinner than those commonly used ($\sim$ 100 nm), aiming for fewer subband occupations. Theory has shown that fewer or single subband regime is preferred for MZM quantization \cite{Prada2012, Brouwer2012ZBP, Loss2013ZBP}. The device was measured in a dilution fridge, base temperature $ T\sim$ 20 mK. The electron $T$ can be below 40 mK. A total bias voltage together with a lock-in excitation was applied to contact N1 with the current $I$ and $dI$ drained from S1. A voltage meter measured $V$ and $dV$ between N2 and S2. This four-terminal set-up can exclude the contributions of contact resistance as a systematic uncertainty. The tunnel gate (TG) and back gate (BG) were used while SG was not well functional and kept grounded. See Fig. S1 for circuit details. 

The tunneling spectroscopy at zero $B$ (Fig. 1b) resolves a hard superconducting gap. The gap size is $\sim$ 0.39 mV due to the thin Al film. Fig. 1c shows the $B_{\text{z}}$ dependence of the gap. The z axis is 8$^\circ$ misaligned with the nanowire. For $B$ along other directions, see Fig. S2. The gap remains hard below 2 T and gradually becomes soft above 3 T. Hard gap at high $B$ is necessary for MZM quantization since a soft gap destroys the quantization by dissipation \cite{LiuDissipation, DasSarma2017QZBP}.

In Fig. 1d we tune the device to a gate setting and find a ZBP near $2e^2/h$ (see Fig. 1e for the `waterfall' plots). $B$ is aligned with the nanowire unless specified. Along this direction, the maximum field allowed is 1.87 T due to the hardware limit. From 1.3 T to 1.85 T, the zero-bias conductance remains within the $5\%$ tolerance bar, see the pink background (from 0.95 to 1.05). The ZBP width, $\sim$ 0.3 mV, is 20 times larger than the thermal width (3.5$k_BT\sim$ 15 $\mu$eV for $T$ = 50 mK). A wide peak is necessary for MZM or quasi-MZM quantization to minimize the effect of thermal broadening. This requires a large tunnel transmission, reflected by the outside gap conductance in Fig. 1e being $\sim$ 10 times larger than that in Fig. 1c. Large transmission results in a finite subgap conductance. Unlike a soft gap, this Andreev reflection induced subgap conductance does not affect the MZM quantization \cite{DasSarma2017QZBP}.

We then set $B$ to 1.35 T and scan gate, see Fig. 1f (for `waterfall' plots, see Fig. S3). The zero-bias conductance remains near $2e^2/h$ over a sizable gate range. Occasionally it slightly deviates from the $5\%$ tolerance bar due to small oscillations. The oscillations in $V_{BG}$, also observed in our previous work \cite{Song2021}, are possibly due to the formation of an open island as visualized by the weak Coulomb blockade diamond in the color map. The island is likely defined between the barrier region and the S2 contact which forms a weak barrier due to the ultra-thin diameter. The oscillation causes a small splitting of the ZBP, suggesting the energy also being modified. 

The three scans in Figs. 1d and 1f were in close measurement sequence without noticeable charge jumps in between. These three scans simultaneously pass through one `sweet spot' in the parameter space ($B$ = 1.35 T, $V_{\text{TG}}$ = -9.137 V, $V_{\text{BG}}$ = 6.886 V), and their zero-bias line cuts all resolve a plateau feature near $2e^2/h$. The plateau in $V_{\text{TG}}$ exceeds 100 mV, significantly wider than a fine-tuned sharp crossing. We therefore identify this combined feature as a ZBP-plateau near $2e^2/h$, mostly within the $5\%$ tolerance bar. 

Fig. 1g shows the zero-bias map by scanning both $B$ and gate voltages, also passing through this `sweet spot'. The three-color plots resolve the `red islands' as the $2e^2/h$-regions (within $\pm 5 \%$). Occasionally, the conductance can slightly exceed 1.05, by $\sim 1\%$ (the cyan region). The `red island' is similar to simulations on partially separated MZMs \cite{Tudor2021Disorder}. Based on the Coulomb blockade diamond size $\sim$ 60 $\mu$V and its period in $V_{\text{BG}}$ ($\sim 12.7$ mV) in Fig. 1f, we can extract the lever arm between $V_{\text{BG}}$ and the energy scale possibly related to the electro-chemical potential. The size of the `red island' in Fig. 1g is $\sim$ 65 mV in $V_{\text{BG}}$, corresponding to an energy scale of $\sim$ 300 $\mu$eV.

In the MZM or quasi-MZM picture, several factors could cause the deviation of ZBP from $2e^2/h$: 1) soft gap; 2) thermal broadening; 3) residual tunnel coupling between the second MZM and the probe; 4) coupling between the two MZMs; 5) multiple subband occupation. The first factor, a soft gap \cite{Takei2013}, not only destroys ZBP quantization, but is also detrimental to MZM applications \cite{Chang2015, Zhang2017Ballistic} and should be avoided. This problem has been solved by the observation of a hard gap within the interested $B$ range (Fig. 1c). The second factor lowers the ZBP height. Since our ZBP width is mainly tunnel broadened and $\sim$ 20 times larger than the thermal width, this effect is also small: thermal averaging at $T$ = 50 mK only causes a height change of $\sim 1\%$ for such a wide peak. The third and fourth factors are closely related and can either increase or decrease the height from $2e^2/h$, depending on the coupling details \cite{Tudor2021Disorder}. For the last factor, a second channel could provide a small additional conductance background, increasing the total height above $2e^2/h$ \cite{Wimmer2011QPC}. Based on its saturation conductance (Fig. S2), our thin nanowire is likely still not in the single subband regime yet. The small zero-bias conductance in Figs. 1d-g (at non-ZBP regions), which can be less than 4$\%$ of $2e^2/h$, suggests that the multi-band contribution to the background conductance (if any) is small.

\begin{figure}[htb]
\includegraphics[width=\columnwidth]{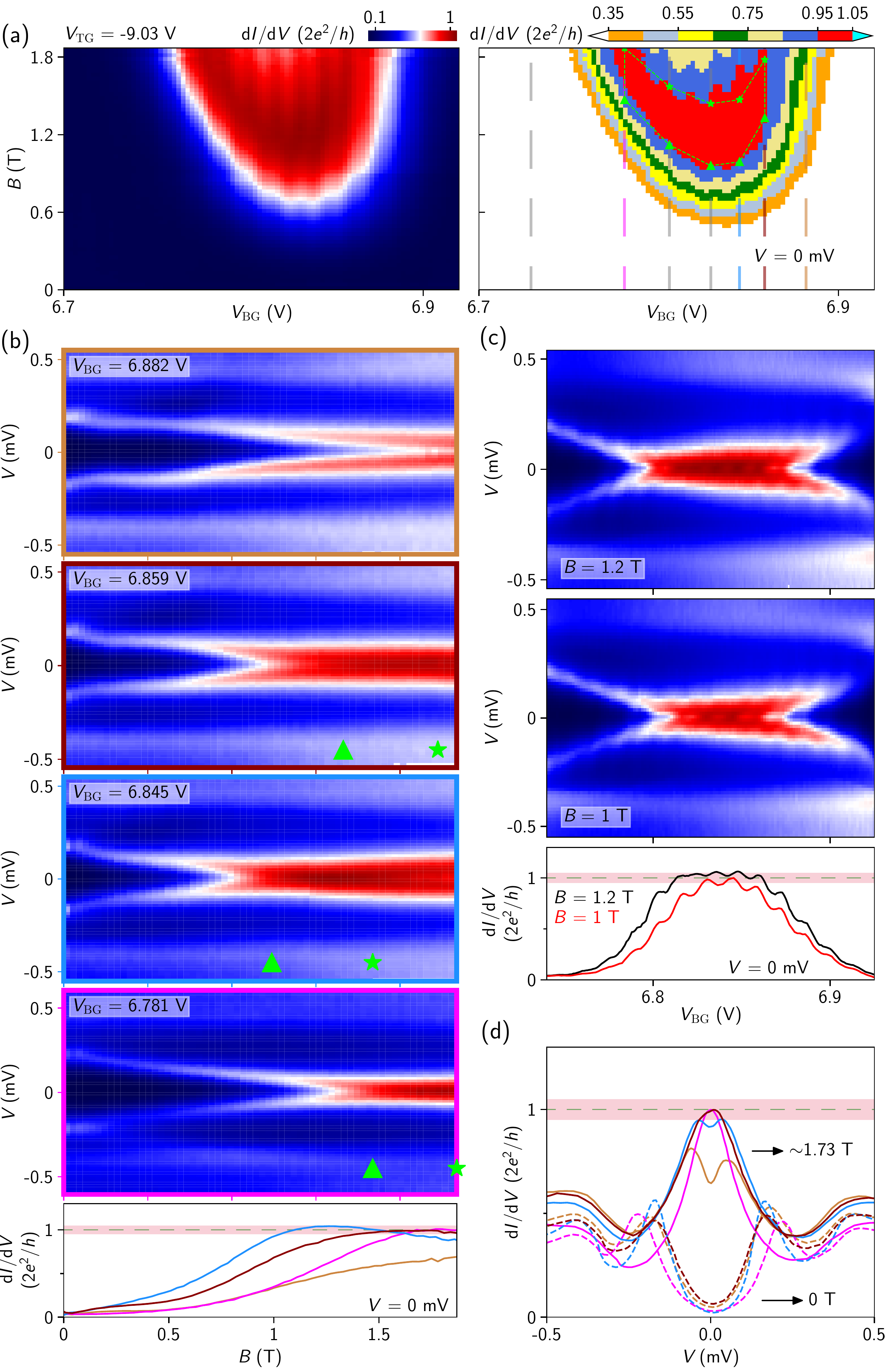}
\centering
\caption{$V_{\text{TG}}$ = -9.03 V for all panels. (a) Zero-bias map in ($B$, $V_{\text{BG}}$) space. The right panel is the multi-color plot. (b) $B$ scans at four $V_{\text{BG}}$ settings, see the dashed lines in (a). The bottom panel shows the zero-bias line cuts, with corresponding colors. (c) $V_{\text{BG}}$ scan at 1.2 T (top) and 1.0 T (middle) with zero-bias line cuts shown in the bottom. (d) Line cuts from (b) at 0 T (dashed) and $\sim$ 1.73 T (solid) with corresponding colors. All panels share the same color bar.}
\label{fig2}
\end{figure}

\begin{figure}[htb]
\includegraphics[width=\columnwidth]{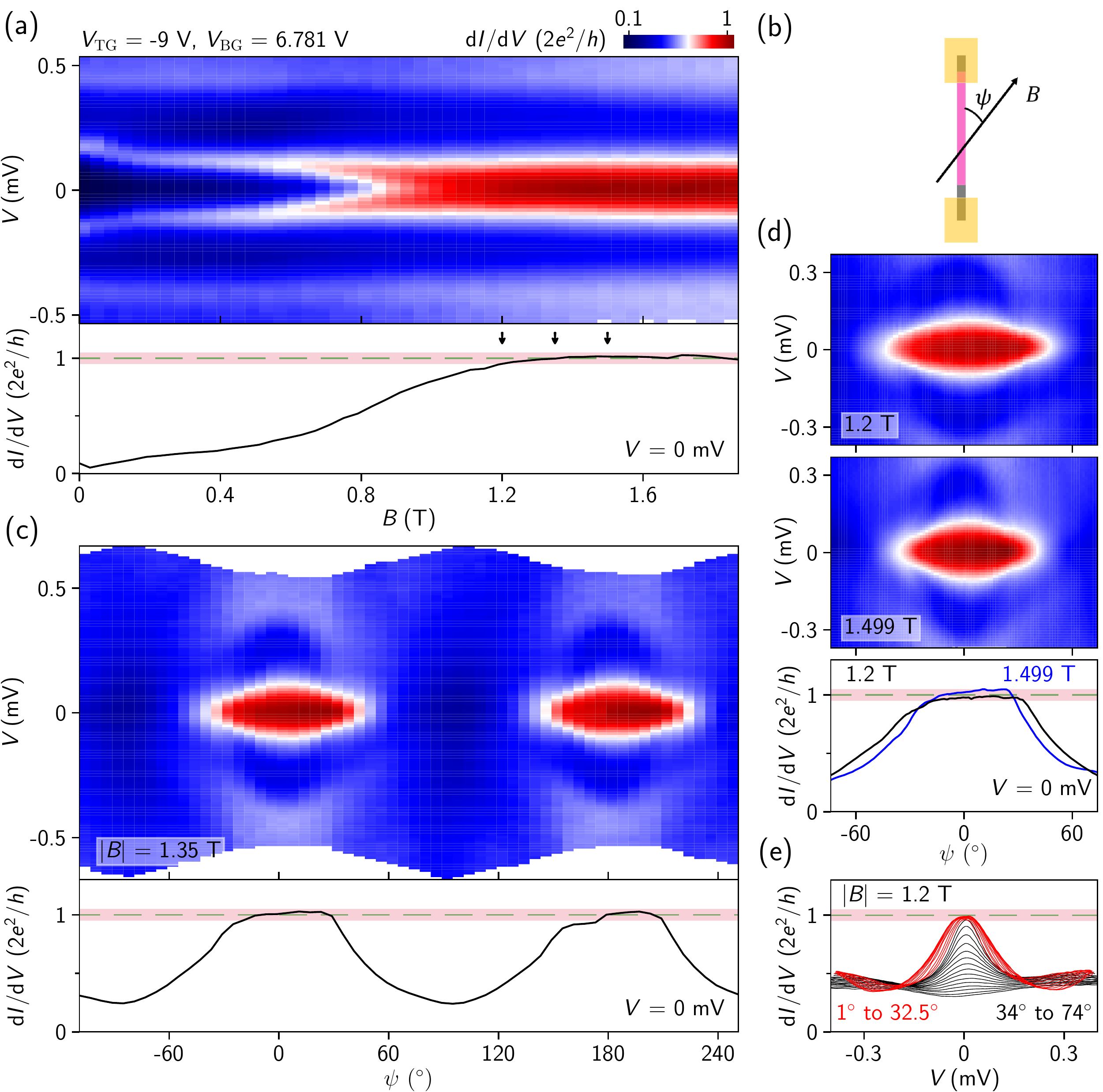}
\centering
\caption{(a) $B$ scan of a ZBP near $2e^2/h$. Lower panel, zero-bias line cut. (b) Schematic illustration of the angle $\psi$. $B$ is in-plane. (c)  Angle dependence of the ZBP by fixing the $B$ amplitude at 1.35 T, see the middle black arrow in (a). Lower panel, zero-bias line cut.  (d) Two more rotation scans at 1.2 T (upper) and 1.499 T (middle), see the other two arrows in (a). Lower panel, zero-bias line cuts. (e) Line-cuts from (d) (upper). For clarity, half of the angle range (and every other line cut) is shown. All panels share the same color bar.}
\label{fig3}
\end{figure}

We next fix $V_{\text{TG}}$ to a slightly different value and explore the zero-bias map in ($B$, $V_{\text{BG}}$) space, see Figure 2a. Besides the `red island' as a `zone', we use several discrete colors to label other ranges which are `boundary-like'. Fig. 2b shows four $B$ scans, see the corresponding colored dashed lines in Fig. 2a. The first scan is outside the `red island' and resolves no ZBPs. Other scans pass through the `red island' and resolve ZBPs near $2e^2/h$ with different $B$ ranges. The green triangles and stars mark the on-set and ending $B$ values for ZBPs. The ZBP region defined by the green dashed lines in Fig. 2a roughly matches the `red island' and can serve as a $2e^2/h$-ZBP phase diagram. Note that not every line cut passing through the `red island' can resolve a plateau feature. For example, the blue line in Fig. 2b is more `peaked-like' near $2e^2/h$.

Fig. 2c shows $V_{\text{BG}}$ scans of the ZBP at 1.2 T and 1 T, corresponding to the middle and edge of the `red island', respectively. The zero-bias line cuts illustrate the evolution from `peak-like' (red)  to `plateau-like' (black). Fig. 2d shows the line cuts from the four panels in Fig. 2b at $B$ = 0 T (dashed lines) and $\sim$1.73 T. At zero field, the subgap peaks have similar energies. At high field (1.73 T), some curves resolve ZBPs near $2e^2/h$ while others resolve split peaks, reflecting the phase diagram boundaries. For `Waterfall' plots and additional scans see Figs. S3 and S4.

The next experimental knob is $B$ direction. $B$ rotation in Al-based systems was limited before due to orbital effects: the Al bulk gap is easily suppressed at low $B$ ($\sim$ 0.3 T) during rotation \cite{Deng2016}. Here, the ultra-thin diameter significantly suppresses the orbital effect and the gap can survive at high $B$ even if misaligned (see Fig. S2). This advantage enables a rotation for ZBP \cite{Jouri2019}.

\begin{figure}[htb]
\includegraphics[width=\columnwidth]{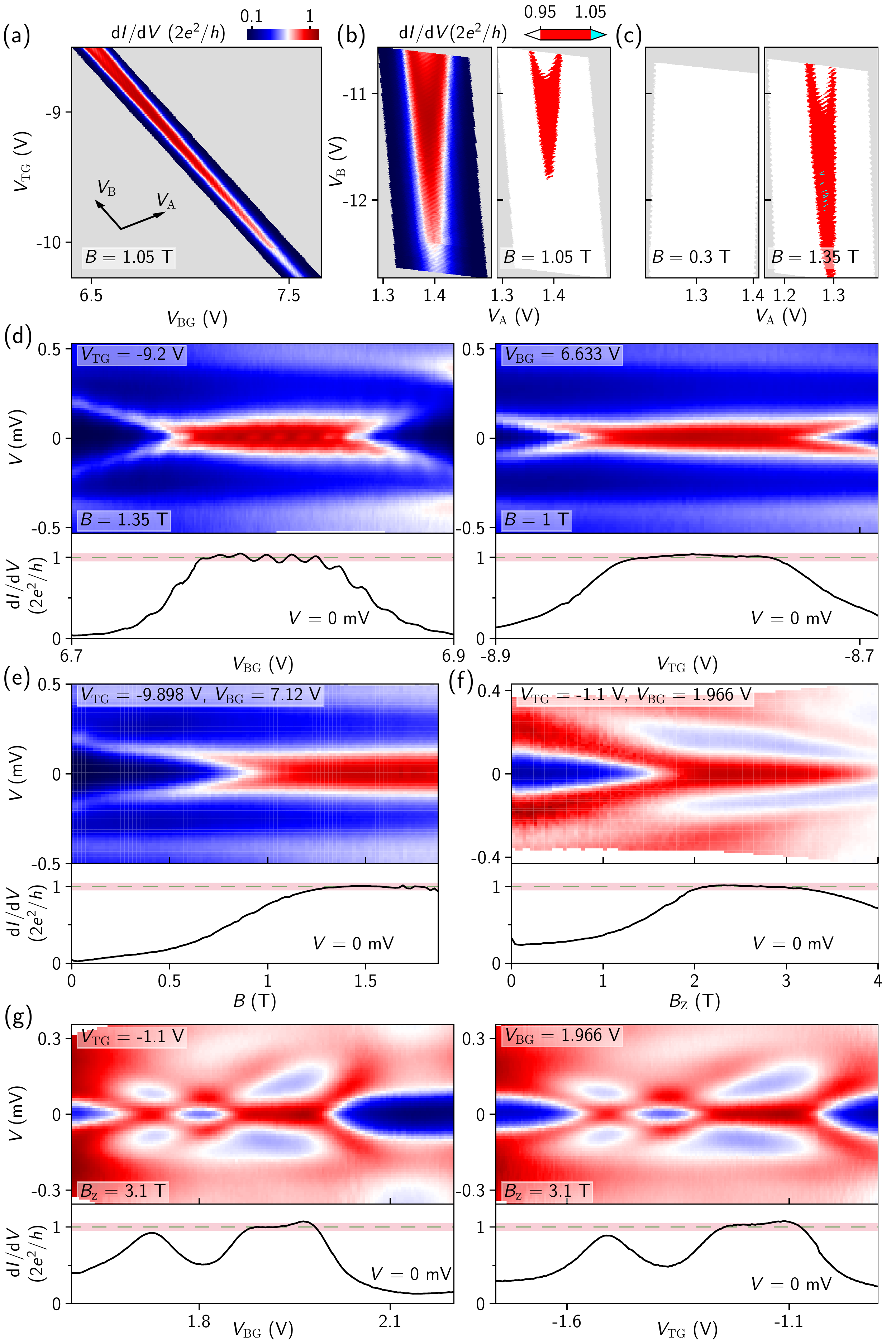}
\centering
\caption{(a) Zero-bias map in ($V_{\text{TG}}$, $V_{\text{BG}}$) space at 1.05 T. Grey regions have no data. (b) Re-plotting (a) using the new axis $V_{\text{A}}$ and $V_{\text{B}}$ (see labeling). Right panel, three-color plot. (c) Zero-bias maps at 0.3 T and 1.35 T in three-color plots. (d-e) Three additional example scans ($V_{\text{BG}}$, $V_{\text{TG}}$, $B$) for ZBPs within region I. (f-g) $B_{\text{z}}$ and gate scans of a ZBP in region II. All panels share the same color bar.}
\label{fig4}
\end{figure}

Figure 3a starts with a $B$ scan (aligned with the nanowire) of a ZBP near $2e^2/h$ (note the minor charge jump at $\sim$ 1.7 T). The $B$ amplitude is then fixed at three values, see the arrows. Its direction, defined by the angle $\psi$ (Fig. 3b), is rotated. $B$ is in-plane during rotation (parallel to the substrate). Fig. 3c shows the angle dependence at 1.35 T. The ZBP remains close to $2e^2/h$ over an angle range of $\sim$ 50$^{\circ}$. Outside this range, the ZBP height quickly decreases away from $2e^2/h$, accompanied by a gap closing. Finally, the ZBP vanishes at larger angles. For $\psi \sim$ 170$^{\circ}$, the zero-bias conductance drops slightly below 0.95$\times 2e^2/h$ due to a small peak-splitting which is separated by a minor charge jump. Fig. 3d shows the angle dependence at two other $B$s, with line cuts shown in Fig. 3e. For more rotation scans, see Fig. S5.

For the last experimental check, we fix $B$ (aligned with the nanowire) at 1.05 T, and measure the zero-bias map in ($V_{\text{TG}}$, $V_{\text{BG}}$) space, see Figure 4a. Due to the cross talks between the two gates, every gate can tune both the tunnel transmission in the barrier region and the electro-chemical potential in the proximitized wire. Therefore, the $V_{\text{BG}}$ range is adjusted simultaneously for different $V_{\text{TG}}$ settings to trace the same feature. To save space from the `no-data' region (grey), we define new gate voltage axes: $V_{\text{A}} = V_{\text{BG}}$ cos$\phi + V_{\text{TG}}$ sin$ \phi$; $V_{\text{B}} = -V_{\text{BG}}$ sin$\phi + V_{\text{TG}}$ cos$ \phi$. $\phi$ = 30$^{\circ}$. Fig. 4b re-plots Fig. 4a in ($V_{\text{A}}$, $V_{\text{B}}$) axes. Together with Fig. 4c,  the three-color plots show the gradual evolution of the `red island': its area increases from none at low $B$ to a sizable zone at high $B$. For a complete evolution, see Fig. S6. Figs. 4d and 4e show three additional examples of ZBP scans within this parameter region. 

So far, all the ZBPs (from Fig. 1d to Fig. 4e) are within one single region (region I) in the multi-dimensional parameter space ($B$, $V_{\text{TG}}$, $V_{\text{BG}}$), see Fig. S7 for additional scans in this region. The measurement in region I lasts for three weeks with several charge jumps in between. We note the effects of jumps and hysteresis mainly shift the gate voltages, maximally by 300 mV. The main plateau features still remain after the shift with minor variations. 

Fig. 4f and 4g demonstrate another plateau region (region II) at a very different ($B$, $V_{\text{TG}}$, $V_{\text{BG}}$) setting from region I. To access higher $B$s, we apply $B$ along the fridge z axis ($B_{\text{z}}$). The small misalignment (8$^{\circ}$) between $B_{\text{z}}$ and the nanowire should not bring a big difference based on the rotation experiment in Fig. 3. In Fig. 4f, the zero-bias conductance can stick close to $2e^2/h$ from 2 T to 3.28 T, a $B$-range larger than 1.2 Tesla (with a tolerance of 5$\%$). Outside this range, the ZBP height drops continuously from $2e^2/h$ with a faster decreasing rate, possibly due to the combined effects of gap closing and softening for $B$ larger than 3 T. The fridge base $T$ for Fig. 4f is slightly higher (can reach $\sim$ 30 mK) which may also play a role. Even within the plateau region, there is still a decreasing trend with a much smaller slope, possibly due to those mechanisms above.

The overall conductance (barrier transmission) in region II is higher than that in region I, leading to a larger subgap conductance. Fig. 4g illustrates the interaction between the zero-energy state and a quantum dot level, possibly formed near the barrier. Tuning the dot level towards zero energy causes the splitting of the ZBP, similar to the behavior of partially separated MZMs \cite{Prada2017, Clarke2017}. We translate the plateau width in $V_{\text{BG}}$ to an energy scale of $\sim$ 260 $\mu$eV based on our extracted lever arm. For additional scans of region II, see Fig. S8. Though the plateau regions (I and II) in gate voltage space is small, only 50 - 100 mV. Its corresponding energy scale, $\sim$ 260 - 300 $\mu$eV, is much larger than the thermal energy.  

We note that this work addresses the question ‘whether ZBP-plateaus near $2e^2/h$ exist by sweeping all relevant experimental parameters’. It, however, cannot answer the question ‘whether similar plateaus (for sweeping all parameters) also exist at non-quantized values’. That would require exhausting the entire parameter space in more devices. Based on the collected data so far, we have not observed similar plateaus at other values yet. See Fig. S9 for additional ZBP data due to quantum dots and disorder in regions beyond I and II. Occasionally, we could find `plateau-like' features above $2e^2/h$ for one-parameter sweeping (but not all). We notice that for multiple subband occupation, even MZM could lead to some `plateau-like' features above $2e^2/h$ in the open barrier regime with slight disorder \cite{Wimmer2011QPC}. Finally, we compare some scans in Fig. S10 to illustrate the interplay between the barrier transmission and the ZBP height.  

To summarize, we have observed ZBPs near $2e^2/h$ which form a plateau mostly within $5\%$ tolerance by sweeping gate voltages and magnetic field. Our result is qualitatively consistent with the (multi-subband) MZM as well as quasi-MZM theories: the first is topological while the second is trivial and caused by non-uniform potential or disorder. The quasi-MZM scenario is probably more likely. This work is from a single device whose quality is an improvement compared to our previous one \cite{Song2021} but still not in the ballistic regime yet due to the obvious presence of quantum dots and disorder. Besides the `end-to-end' correlation experiment to reveal gap-closing-reopening \cite{Andreev_rectifier, DasSarma_nonlocal, Pikulin_protocol}, future devices on `single-terminal' experiments could aim at 1) looking for zero-bias peak-to-dip transition near $2e^2/h$ with better tolerance \cite{NextSteps,WimmerQuasi, Song2021}; 2) using a dissipative probe to reveal quantized ZBPs and suppress others \cite{Dong_PRL2013,Dong_PRB2020, Dong_2021, ZhangShan, WangZhichuan}; 3) exploring potentially better material systems \cite{CaoZhanPbTe,Jiangyuying}.

\section{Acknowledgment} 

We thank Leo Kouwenhoven for comments. Raw data is available at https://doi.org/10.5281/zenodo.6546974. This work is supported by Tsinghua University Initiative Scientific Research Program, Alibaba Innovative Research Program, National Natural Science Foundation of China (Grant Nos. 12104053, 92065106, 92065206, 61974138, 12004040, 11974198). D.P. also acknowledges the support from Youth Innovation Promotion Association, Chinese Academy of Sciences (Nos. 2017156 and Y2021043).

\bibliography{mybibfile}

\newpage
\onecolumngrid
\newpage



\section*{Supplementary material}

# Supplementary Information for "Plateau Regions for Zero-Bias Peaks within 5% of the Quantized Conductance Value $2e^2/h$"

Zhaoyu Wang, Huading Song, Dong Pan, Zitong Zhang, Wentao Miao, Ruidong Li, Zhan Cao, Gu Zhang, Lei Liu, Lianjun Wen, Ran Zhuo, Dong E. Liu, Ke He, Runan Shang, Jianhua Zhao, and Hao Zhang

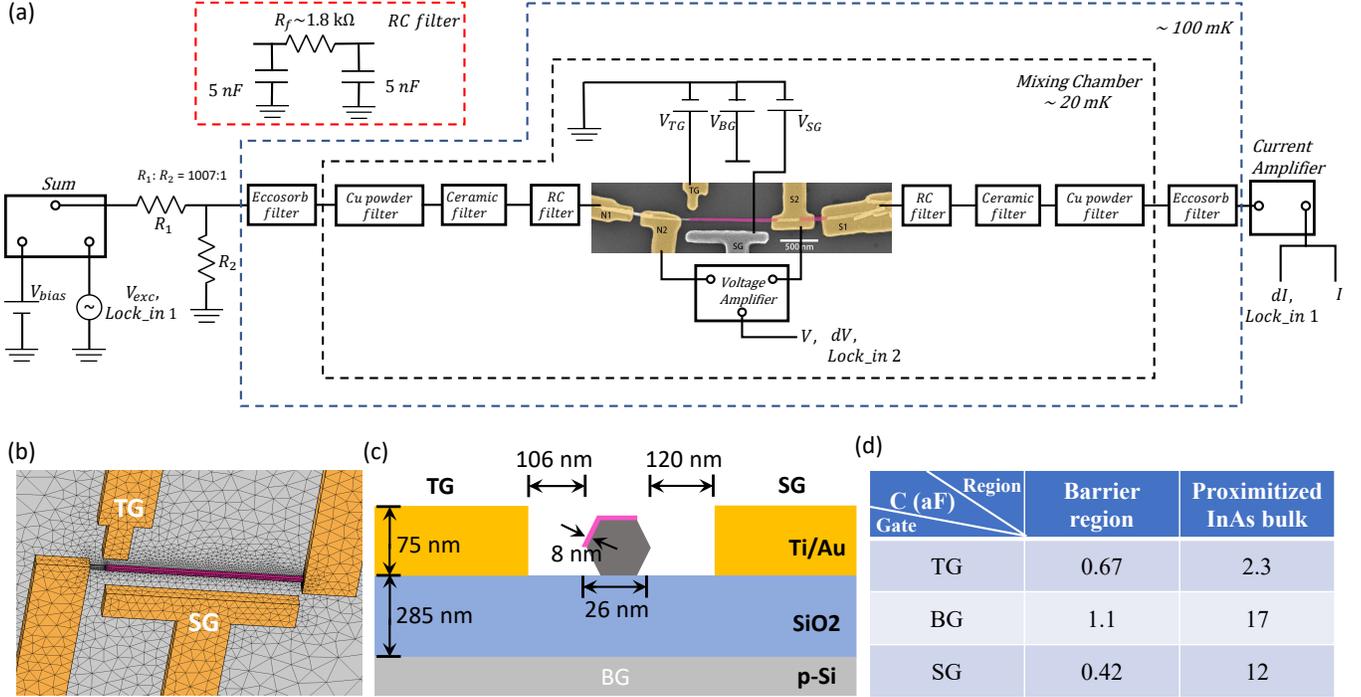

Fig. S 1: (a) Device measurement setup. The InAs-Al nanowire was first transferred onto a p-doped Si/SiO$_2$ substrate. Part of the Al film was then selectively etched using Transene Aluminum Etchant Type D at 50 °C for 8.5 seconds. Contacts and gates were deposited, after a short Argon plasma etching, by the evaporation of Ti/Au (5/70 nm). Transport was measured in an Oxford dilution refrigerator equipped with a 6-3-1.5 T vector magnet. Each fridge line has an eccosorb filter, a copper powder filter, ceramic filters (LFCN-1000 and 3000) and an RC filter (not shown for the gate lines in the schematic). The electron $T$ has been calibrated using a GaAs/AlGaAs quantum Hall thermometer, revealing an electron $T$ lower than 40 mK. During the measurement, a total bias voltage $V_{\text{bias}}$ was first summed with a lock-in excitation, and then applied to the N1 contact after passing through a voltage divider and the four-stage filters. The current $I$ and $dI$ were measured from S1 contact. A voltage amplifier was used to measure $V$ and $dV$ between N2 and S2. The device conductance $G$ as a function of $V$ was directly calculated by $dI/dV$. For $V$, we calibrate the zero bias by subtracting a small offset $V_{\text{offset}}$, estimated from the data set based on its particle-hole symmetry. The fridge line of SG is probably broken, therefore not working. (b-d) Electrostatic simulation for estimating the gate-nanowire capacitance using finite element analysis. (b) and (c) are the device geometry and (d) is the estimated capacitance matrix. While both TG and BG have significant capacitance coupling to the barrier region as reflected by the strong cross talk shown in Fig. 4a, BG also has a much stronger capacitance coupling to the proximitized nanowire bulk (even more than the SG if being functional). This suggests that both the barrier region and the proximitized bulk could be tuned by BG. In addition, the ratio of the plateau width in $V_{\text{TG}}$ vs in $V_{\text{BG}}$ roughly matches the capacitance ratio of BG and TG on the barrier region. This suggests that tuning the barrier segment can have a significant effect on ZBPs. Though the picture of MZMs in a finite wire coupled to a quantum dot might be consistent with our data, a more likely and natural explanation is probably the quasi-MZMs formed near the barrier. As for the small zero-bias oscillations in Fig. 1f, a feature possibly arising from the proximitized bulk, the ratio of $V_{\text{BG}}$ vs $V_{\text{TG}}$ also roughly matches the capacitance ratio of TG and BG on the proximitized nanowire region.

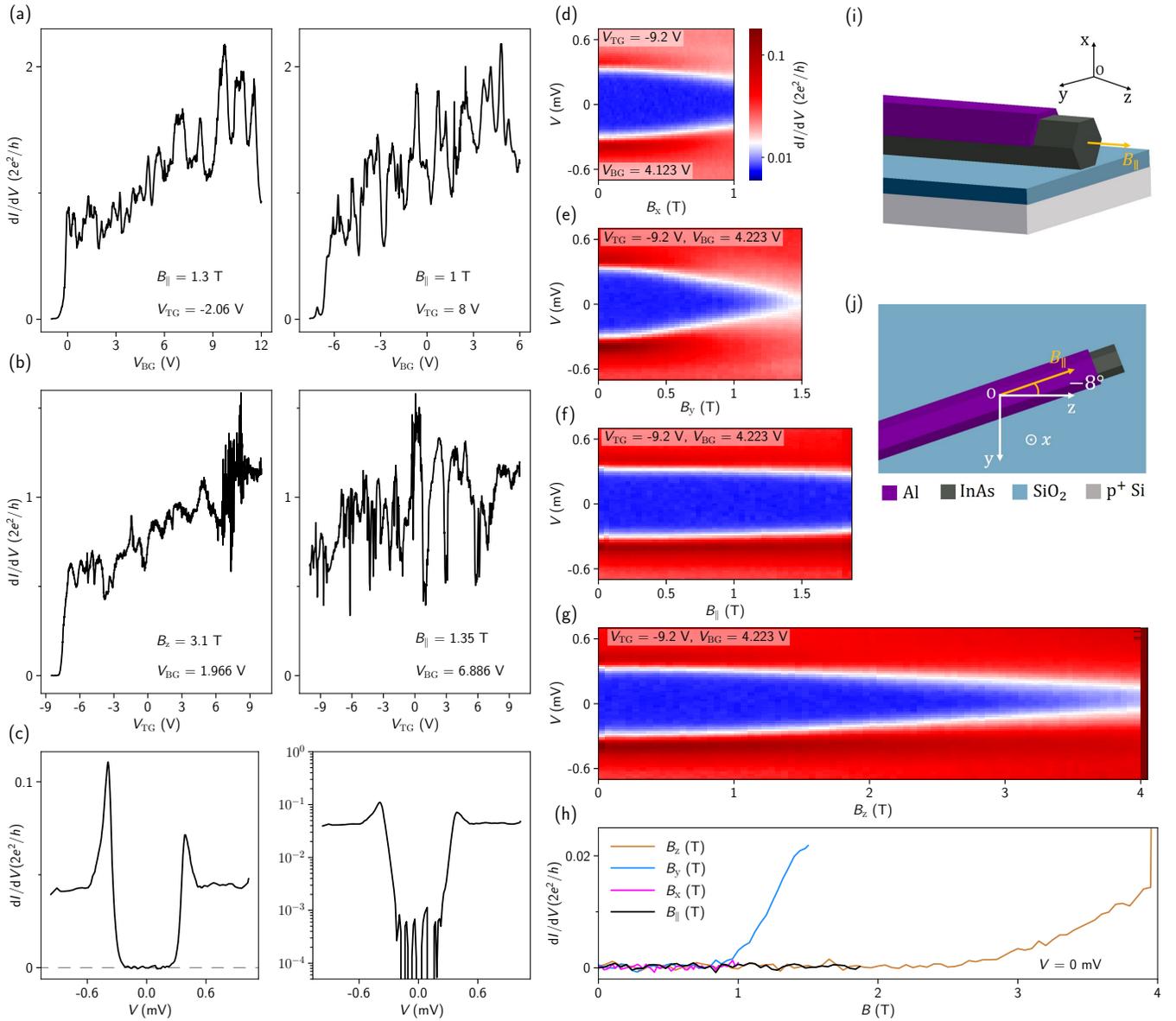

Fig. S 2: (a) $V_{BG}$ pinch off calibration for two different $V_{TG}$ settings. The total bias voltage is fixed at 2 mV and 2.2 mV, respectively. (b) $V_{TG}$ scans with the total bias voltage fixed at 2.5 mV, reflecting the normal state conductance. The normal state conductance larger than $2e^2/h$ is a possible indication of multiple subband occupations. (c) Line cut from Fig. 1b in the tunneling regime in linear and logarithmic scale, resolving a hard superconducting gap. The outside-gap/subgap conductance suppression can reach two orders of magnitude. (d-g) $B$ scans of the hard gap along different directions. (h) Zero-bias line cuts from (d-g). (i) Schematic of the device orientation. The fridge magnetic axis (x, y, z) is labeled. (j) Top view of the device, the nanowire is 8° misaligned with the fridge z axis. (d-g) share the same color bar.

2

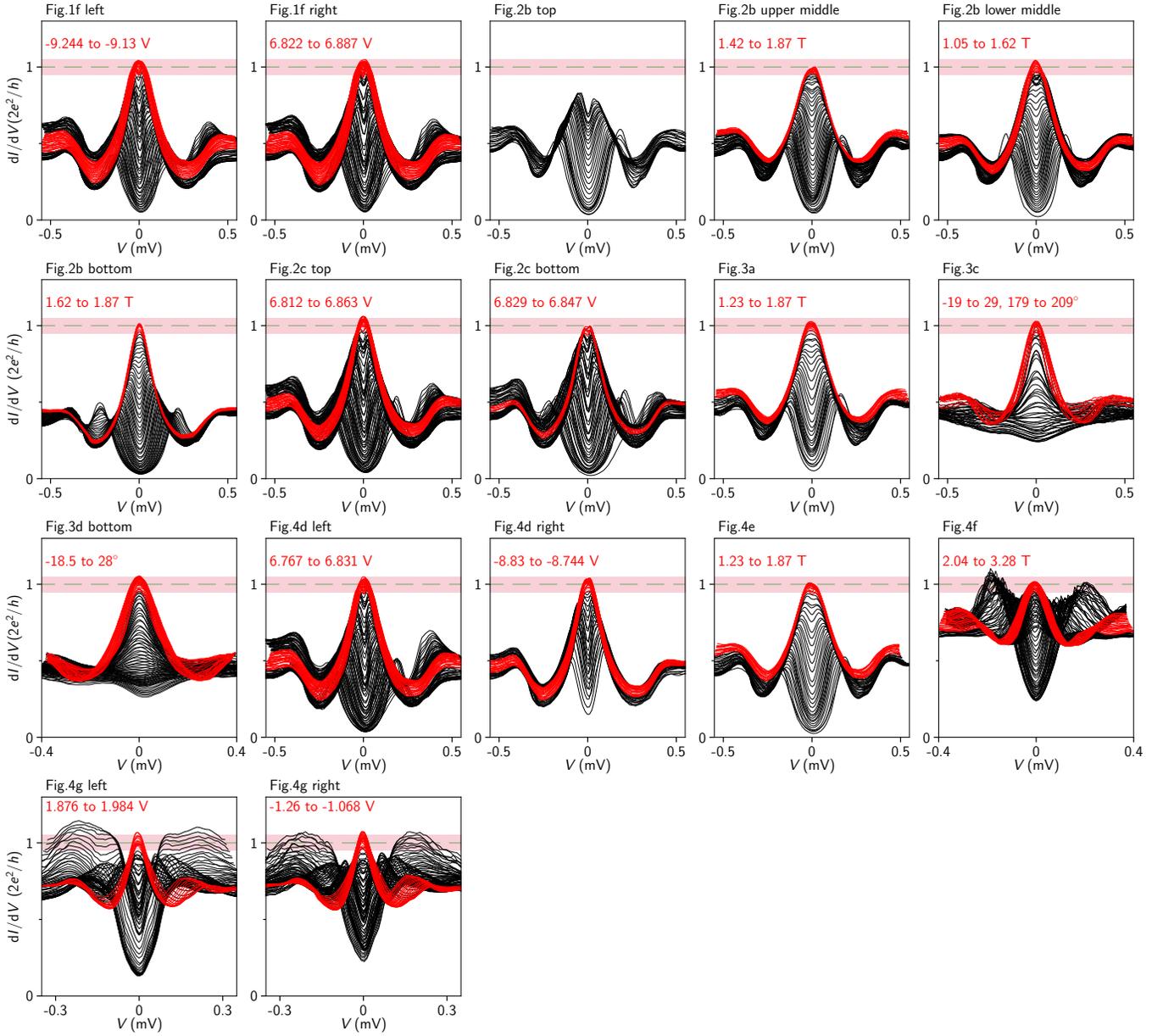

Fig. S 3: 'Waterfall' plots of maintext figures (labeled on top of each panel). For clarity, every other curve is shown in Fig. 1f, Fig. 2b (top), Fig. 2c, Fig. 4d, Fig. 4e and Fig. 4g. Red and black colors are used (also for clarity), see the red labeling for the corresponding parameter range.

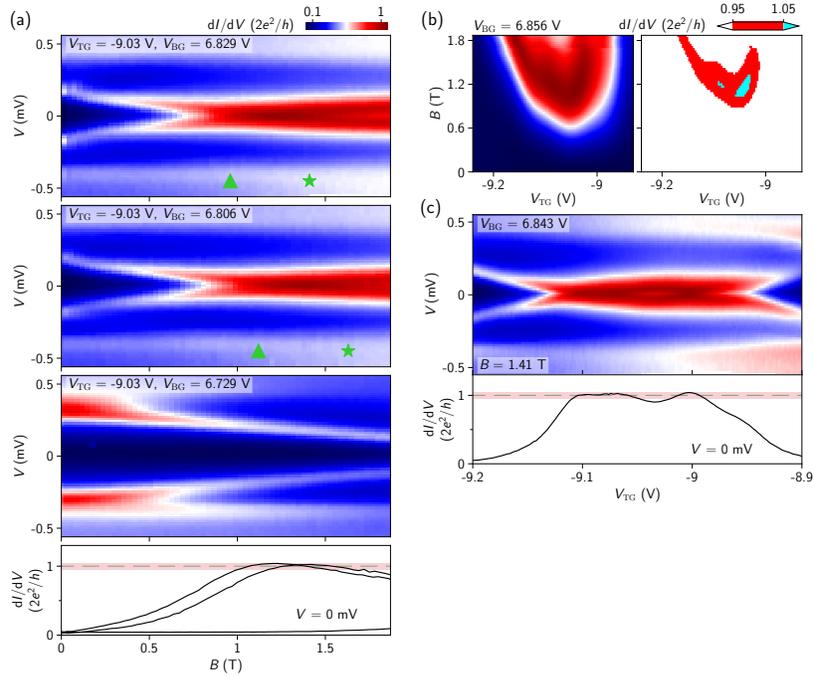

Fig. S 4: Additional scans for Fig. 2. (a) Three $B$ scans, corresponding to the gray dashed lines in Fig. 2a. Lower panel, zero-bias line cuts. (b) Zero-bias map in ($B$, $V_{TG}$) space and its three-color plot. (c) $V_{TG}$ scan of the ZBP. All color-plots share the same color bar.

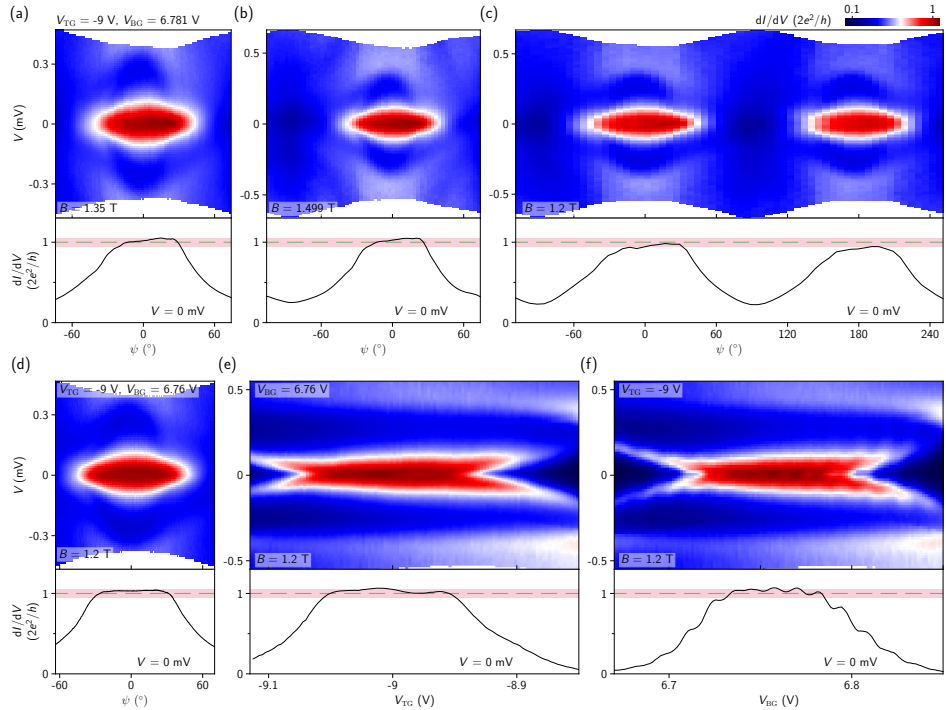

Fig. S 5: Additional scans for Fig. 3. (a-c) Angle dependence of the ZBP at 1.35 T, 1.499 T and 1.2 T. These measurements repeat the ones in Fig. 3c and 3d but with different angle ranges. (d) Angle dependence of the ZBP at a slightly different gate setting. (e-f) Gate dependence of this ZBP. All color-plots share the same color bar.

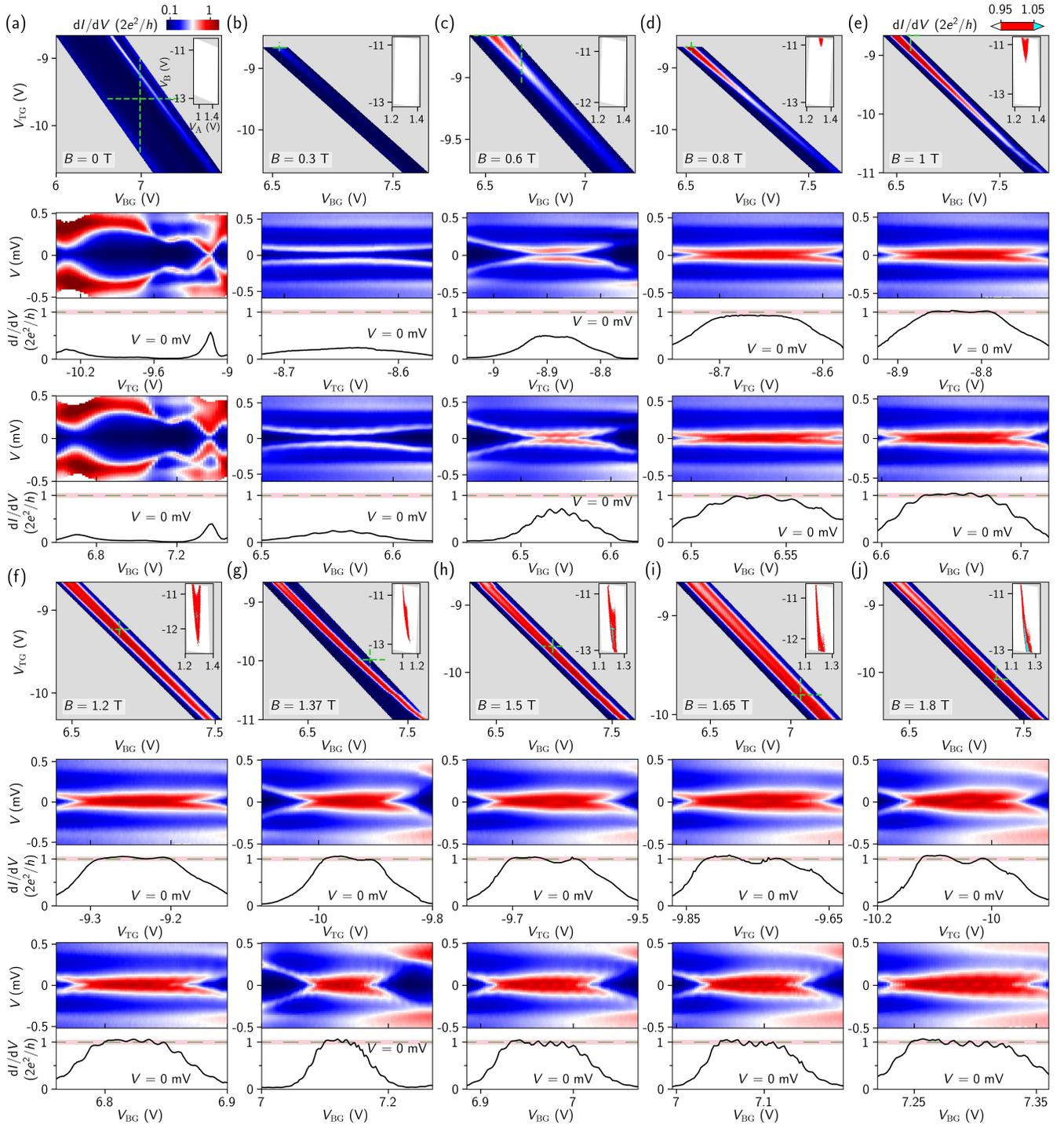

Fig. S 6: (a)-(j) are for $B$ from 0 to 1.8 T to illustrate the continuous evolution of the ZBPs in region I. For each $B$, the top panel is the zero-bias conductance map in ($V_{TG}$, $V_{BG}$) phase space. The gray regions have no experimental data. Inset, three-color plot using the new axes ($V_A$, $V_B$). The middle and lower panels are $V_{TG}$ and $V_{BG}$ scans with the zero-bias line cuts shown at the bottom. These scans correspond to the dashed green lines in the top panel. Note that there are charge jumps or hysteresis between different scans which reset or shift the gate voltages. All color maps share the same color bar.

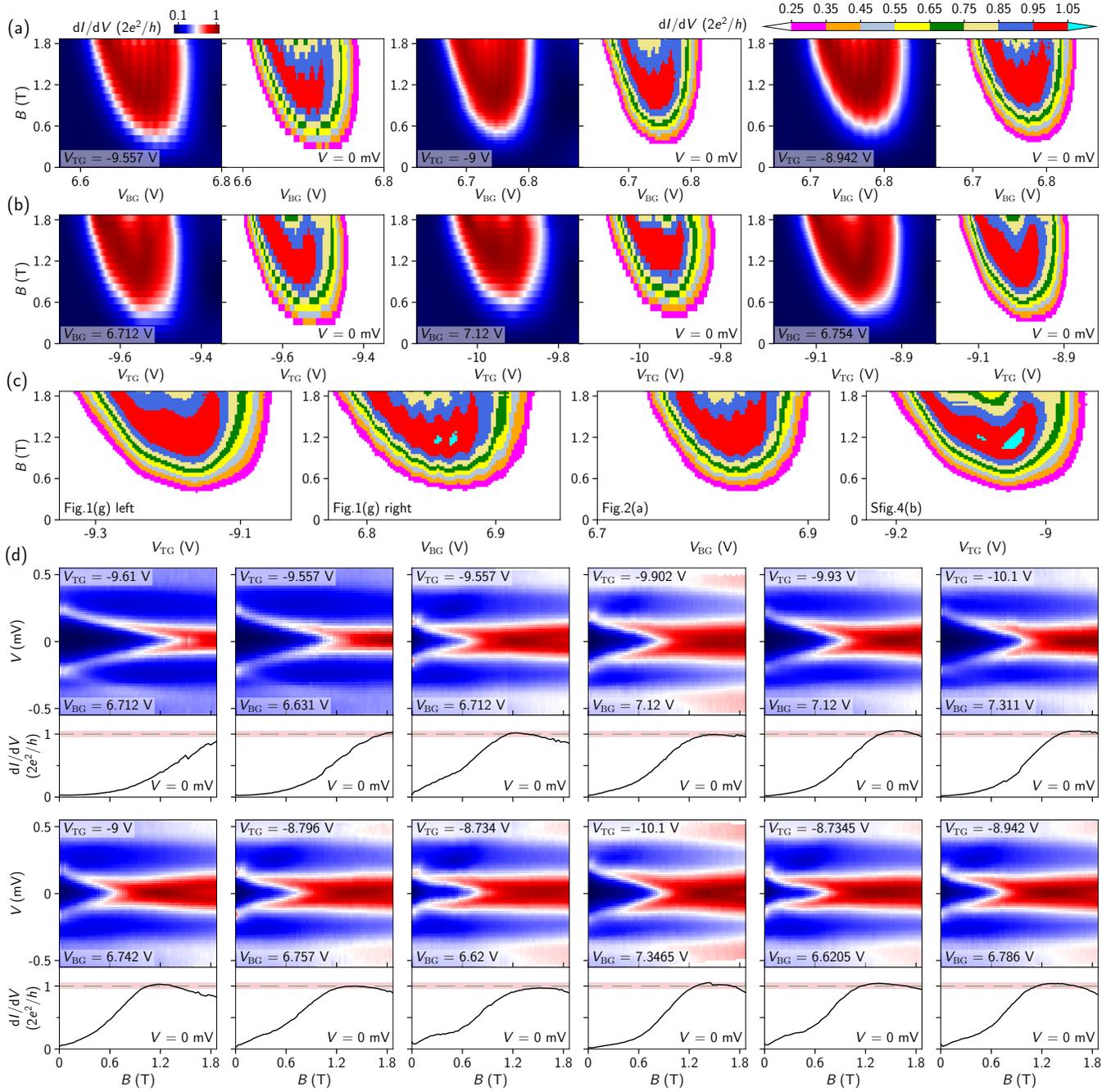

Fig. S 7: Additional various scans for region I. (a) Zero-bias maps in $(B, V_{\text{BG}})$ space and their multi-discrete-color plots for different $V_{\text{TG}}$s. (b) Zero-bias maps in $(B, V_{\text{TG}})$ space and their multi-discrete-color plots for different $V_{\text{BG}}$s. (c) Multi-discrete-color plots for other $B$ and gate scan panels (see labeling). (d) 12 additional $B$ scans of ZBPs with gate voltages labeled. All color-plots share the same color bar.

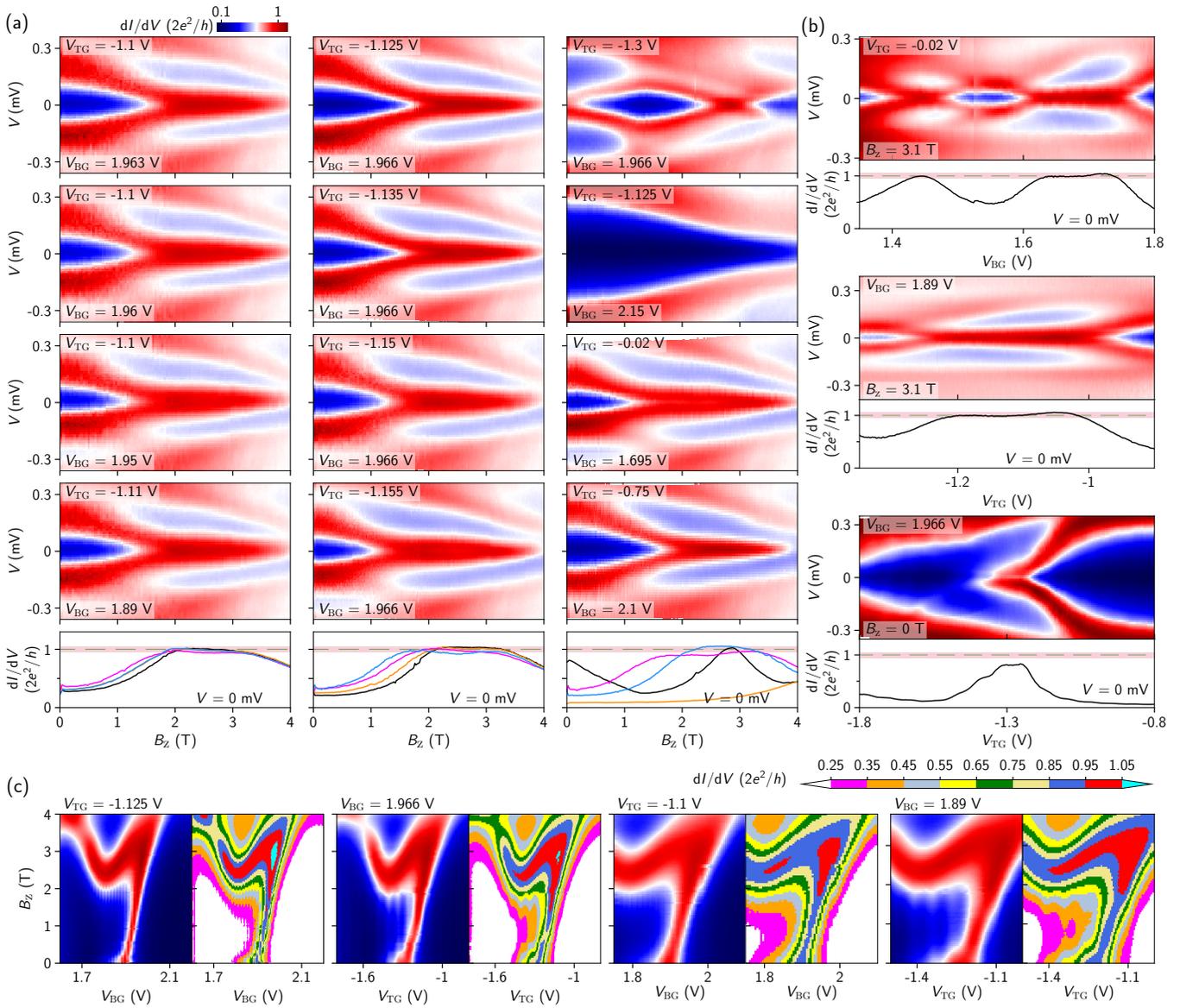

Fig. S 8: Additional scans for region II. (a) 12 additional $B$ scans and their corresponding zero-bias line cuts (bottom panels) within or near region II. Some resolve ZBPs near $2e^2/h$, and some do not (e.g. the top two panels in the third column). The colors for zero-bias line cuts are (from top to bottom) black, orange, pink and blue, respectively. The blue curve in the second column and the pink one in the third column correspond to a split-peak instead of ZBP for their plateaulike feature. (b) $V_{BG}$ (top) and $V_{TG}$ (middle) scans of the ZBP at 3.1 T. The lower panel shows the $V_{TG}$ scan at zero field. Two Andreev bound state (ABS) levels move close to zero at $\sim 1.3$ V, where the corresponding $B$ scan is shown in the top right panel of (a). A typical ABS behavior (at low field) coexists with a ZBP (at high field). The measurement of the top panel in (b) and the lower two panels in the third column of (a) were performed separately from the region II measurements. In between, there were several gate scans over large ranges (See Fig. S2b and Fig. S9j). This results in a gate voltage shift compared to the other scans in region II. (c) Four zero-bias conductance maps in $(B, V_{BG})$ and $(B, V_{TG})$ phase space, with their multi-discrete-color plots. The 'red peak' at zero field corresponds to the ABS level (b) (lower) and (a) (top right). All color-plots share the same color bar.

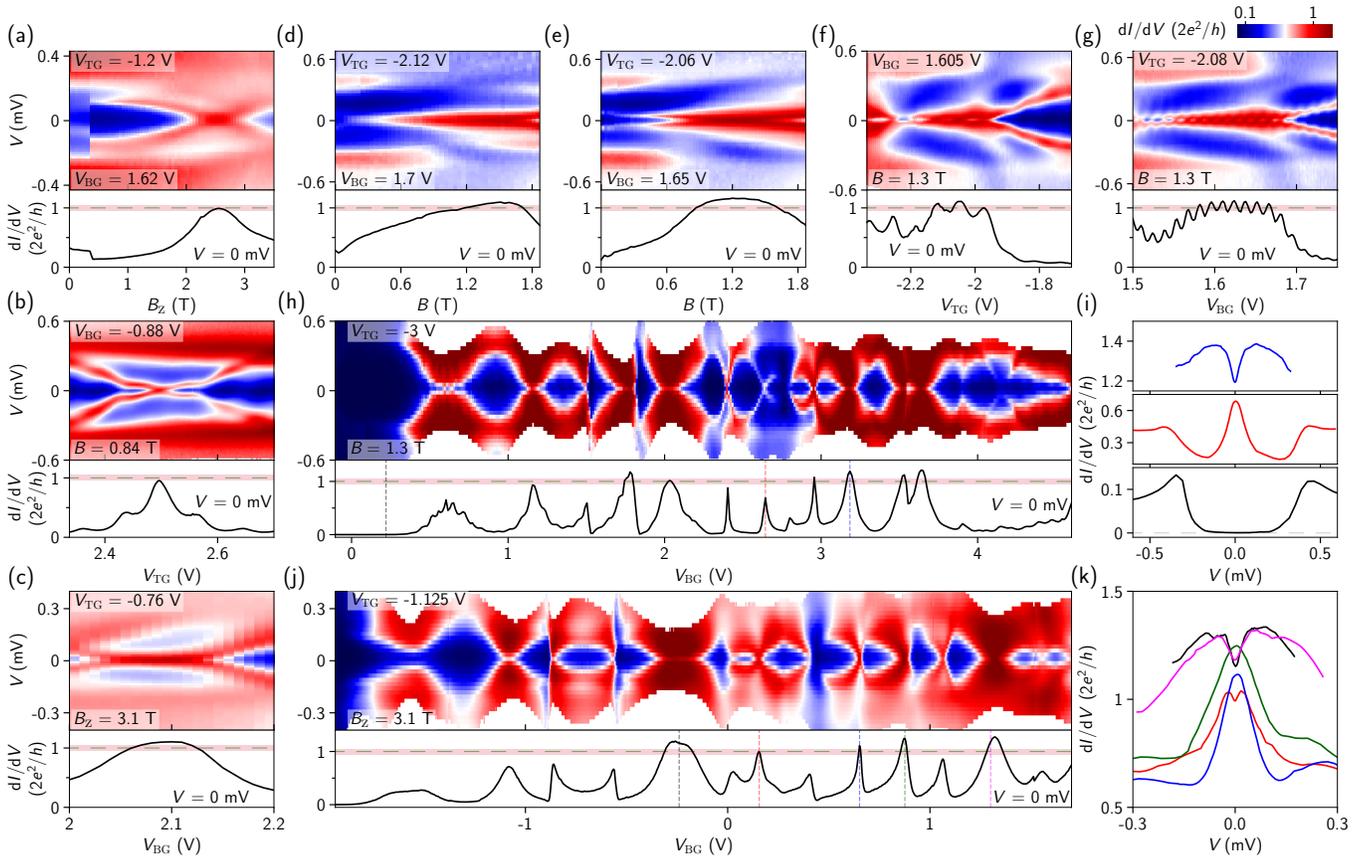

Fig. S 9: Example scans of non-quantized ZBPs beyond region I and II. (a-c) $B$, $V_{TG}$ and $V_{BG}$ scans, all resolving a ZBP whose zero-bias line cut (lower panel) is peaklike instead of plateaulike. The ZBPs in (a) and (b) are near $2e^2/h$ by coincidence. (d) A ZBP remains non-split over a large $B$ range (from 0.15 T to 1.63 T) but its height (zero-bias line cut) does not resolve a plateau. (e) $B$ scan of another ZBP whose zero-bias line cut resolves a plateaulike feature above $2e^2/h$ (outside the 5% tolerance bar). However, the following gate scans on this plateaulike feature in (f) and (g) do not resolve a plateau. Instead, strong oscillations near $2e^2/h$ with amplitudes larger than the 5% bar are revealed. Therefore, (e-g) combined does not meet our definition of plateau. (h) $V_{BG}$ scan over a large gate voltage range at 1.3 T. Lower panel, zero-bias line cut. A non-quantized ZBP (red dashed line) can be resolved as a level crossing, see (i) for the red line cut. The blue dashed line corresponds to a gate-scanned peak, but in a bias scan, it does not resolve a ZBP (see the blue line cut in (i)). (j) Another large range $V_{BG}$ scan at 3.1 T. Lower panel, zero-bias line cut. The black dashed line mark a plateaulike feature, which does not resolve a ZBP in (k). The blue and green line cuts are ZBPs near or above $2e^2/h$, both not resolving a plateau in their gate scan. Finally, we note that there is no clear boundaries between these ZBPs and the ones in the maintext: one might smoothly evolve to the other if the relevant parameter is tuned. All the panels share the same color bar.

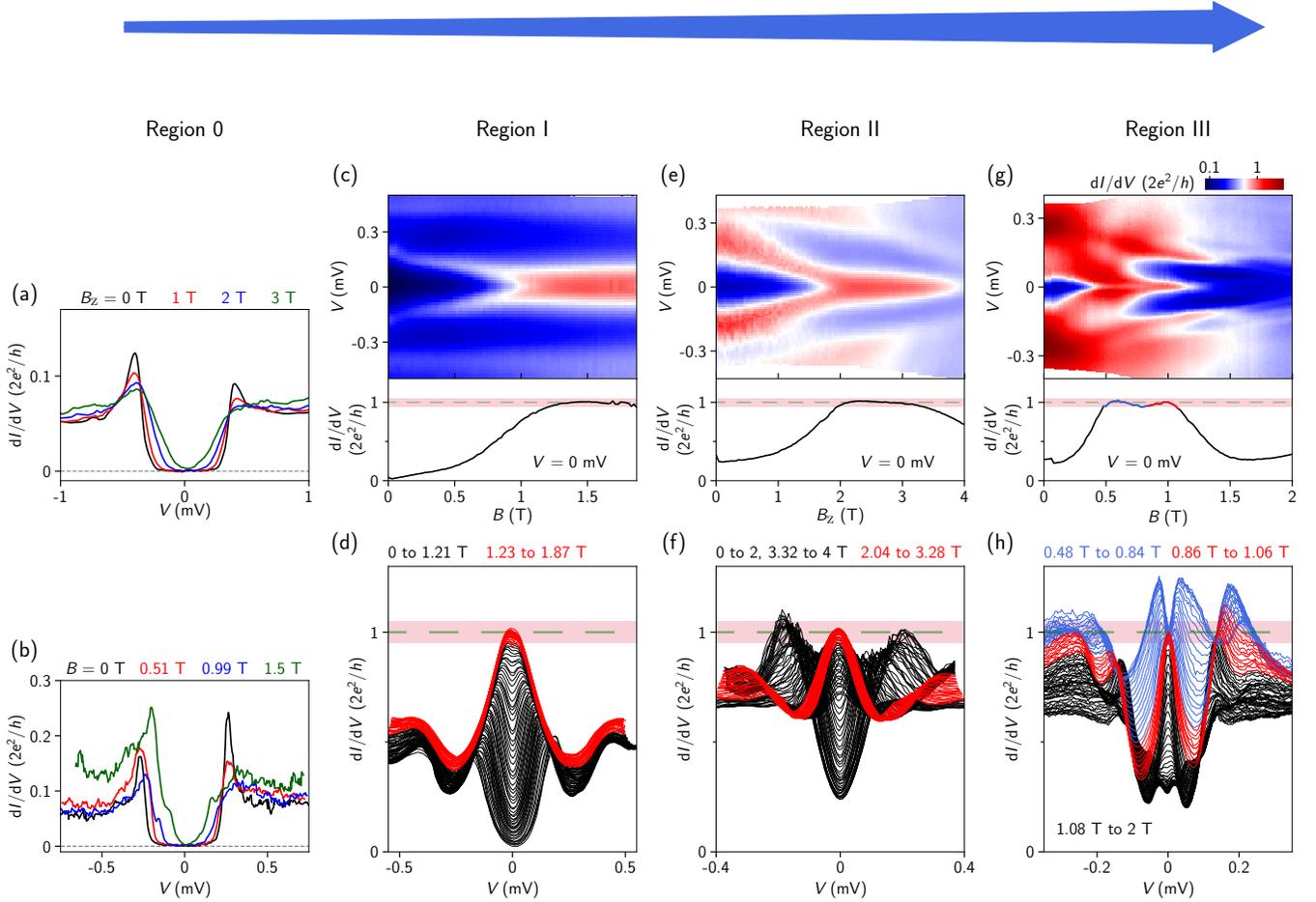

Fig. S 10: Panels (a) and (c-f) are re-plotted from Fig. 1c and Fig. 4e-f. Panels (b) and (g-h) are re-plotted from another four-terminal InAs-Al device in our previous work, Phys. Rev. Research 4, 033235 (2022): here we re-used Fig. 2a and Fig. 11e in that work for illustration purpose. (c), (e) and (g) are color maps (using the same color bar) with the zero-bias line cuts shown in the bottom, while (d), (f) and (h) are the corresponding waterfall plots. From left to right, the barrier transmission increases, as reflected by the outside-gap conductance. In the tunneling regime (region 0, very low transmission), $dI/dV$ relfects the density of states and reveals a hard gap. The gap can remain hard at finite $B$. As the barrier transmission gets higher (from region I to III), large and wide ZBPs can be revealed at finite $B$ near $2e^2/h$ and finally zero-bias dips near $2e^2/h$ can show up (in region III). The ZBP width being much larger than the thermal width ($3.5k_\mathrm{B}T$) is crucial to guarantee the negligible thermal averaging effect.



\end{document}